\documentclass[preprint,12pt,authoryear]{elsarticle}

\usepackage{amssymb}

\usepackage{booktabs}
\usepackage{color}
\usepackage{dcolumn}
\newcolumntype{d}{D{.}{.}{2}}
\usepackage{array}
\newcolumntype{C}[1]{>{\centering\arraybackslash}p{#1}}
\usepackage{url}
\usepackage{color}

\usepackage{multirow}

\journal{NeuroImage}

\begin{document}

\begin{frontmatter}



\title{Choosing Wavelet Methods, Filters, and Lengths for Functional Brain Network Construction}

\author[label1,label2]{Zitong Zhang}
\address[label1]{Department of Biomedical Engineering, Tsinghua University, Beijing 100084, China}
\address[label2]{Department of Bioengineering, University of Pennsylvania, Philadelphia, PA 19104, USA}

\author[label2]{Qawi K. Telesford}

\author[label2,label3]{Chad Giusti}
\address[label3]{Warren Center for Network and Data Sciences, University of Pennsylvania, PA 19104, USA}

\author[label4]{Kelvin O. Lim}
\address[label4]{Department of Psychiatry, University of Minnesota, Minneapolis, MN 55455, USA}

\author[label2,label5]{Danielle S. Bassett}
\address[label5]{Department of Electrical and Systems Engineering, University of Pennsylvania, Philadelphia, PA 19104, USA}


\end{frontmatter}

\newpage
\clearpage

\noindent \textbf{Abstract} Wavelet methods are widely used to decompose fMRI, EEG, or MEG signals into time series representing neurophysiological activity in fixed frequency bands. Using these time series, one can estimate frequency-band specific functional connectivity between sensors or regions of interest, and thereby construct functional brain networks that can be examined from a graph theoretic perspective. Despite their common use, however, practical guidelines for the choice of wavelet method, filter, and length have remained largely undelineated. Here, we explicitly explore the effects of wavelet method (MODWT vs. DWT), wavelet filter (Daubechies Extremal Phase, Daubechies Least Asymmetric, and Coiflet families), and wavelet length (2 to 24) -- each essential parameters in wavelet-based methods -- on the estimated values of network diagnostics and in their sensitivity to alterations in psychiatric disease. We observe that the MODWT method produces less variable estimates than the DWT method. We also observe that the length of the wavelet filter chosen has a greater impact on the estimated values of network diagnostics than the type of wavelet chosen. Furthermore, wavelet length impacts the sensitivity of the method to detect differences between health and disease and tunes classification accuracy. Collectively, our results suggest that the choice of wavelet method and length significantly alters the reliability and sensitivity of these methods in estimating values of network diagnostics drawn from graph theory. They furthermore demonstrate the importance of reporting the choices utilized in neuroimaging studies and support the utility of exploring wavelet parameters to maximize classification accuracy in the development of biomarkers of psychiatric disease and neurological disorders.


\newpage
\clearpage
\section{Introduction}

The use of functional neuroimaging has gained considerable popularity over the last two decades as it provides a noninvasive approach for studying the brain \citep{biswal1995functional}. Although a relatively recent addition to the methods available for analyzing neuroimaging data, network science has enhanced our understanding of the brain as a complex system. Rooted in techniques derived from graph theory, brain network analysis has been used to study neural diseases \citep{boersma2013disrupted}, aging \citep{geerligs2014reduced}, and cognitive function \citep{mantzaris2013dynamic}. The graph theory formalism defines a network by nodes (brain regions) and edges (connections between brain regions). In neuroimaging studies, nodes can describe atlas-based regions \citep{tzourio2002automated} or voxels \citep{eguiluz2005scale,van2008small}, and edges can define physical connections, in the case of anatomical networks \citep{sporns2002theoretical,sporns2007identification,hagmann2008mapping}, or functional connections, which describe a statistical relationship between the activity time series of two nodes \citep{honey2007network,honey2009predicting}.

The goal of generating a brain network is straightforward: to use network science to understand the structure and function of the brain. Most studies report basic graph diagnostics, which include features of individual nodes (e.g., node centralities), features of groups of nodes (e.g., community structure or modularity), or features of the whole brain (e.g., global efficiency). Network analysis can also be used to explore fundamental principles of brain network organization, including small-world architecture \citep{humphries2008network}, cost-efficiency \citep{Bullmore2012}, and reconfiguration dynamics \citep{Bassett2011,Hutchison2013}. Across these studies, the main focus is to understand the organization of nodes and edges in the network. However, what has received less attention is the methodology used to define the functional relationships between nodes. In the context of functional brain networks, popular methods to define statistical relationships between regional activity time series include Pearson's correlation coefficient \citep{Zalesky2012}, coherence \citep{Peters2013}, wavelet correlation \citep{Achard2006}, and wavelet coherence \citep{Bassett2011,Bassett2013a,Bassett2013b,mantzaris2013dynamic,Bassett2014,Bassett2015}; a less common method is the cross-sample entropy \citep{pritchard2014functional}.

Wavelet-based methods have significant advantages in terms of denoising \citep{Fadili2004}, robustness to outliers \citep{Achard2006}, and utility in null model construction \citep{Breakspear2004}. Moreover, wavelet-based methods facilitate the examination of neurocognitive processes at different temporal scales without the edge effects in frequency space that accompany traditional band pass filters \citep{percival2000wavelet}. But perhaps the most compelling argument in support of wavelets \citep{Achard2007} derives from the fact that cortical fMRI time series display slowly decaying positive autocorrelation functions (also known as long memory) \citep{Maxim2005,Wink2006}. This feature undermines the utility of measuring functional connectivity between a pair of regional time series using a correlation (time domain) or coherence (frequency domain), because both time- and frequency-domain measures of association are not properly estimable for long memory processes \citep{Beran1994}. In contrast, wavelet-based methods provide reliable estimates of correlation between long memory time series \citep{Whitcher2000,Gencay2001} derived from fMRI data \citep{Achard2007,Bullmore2004,Achard2008}. Based on these advantages, wavelet-based estimates of functional connectivity have provided extensive insights into brain network organization in health \citep{achard2006resilient}, aging \citep{achard2007efficiency}, neurological disorders \citep{supekar2008network}, sleep \citep{spoormaker2010development}, and cognitive performance \citep{giessing2013human}.

Despite the utility of wavelet-based approaches for estimating functional connectivity, fundamental principles to guide the performance of wavelet-based methods remain largely undefined. This lack of guidelines is apparent in the wide range of wavelet methods, filters, and lengths utilized in graph theoretical neuroimaging studies, which hampers comparability and reproducibility of subsequent findings. Here we explore the use of different wavelet methods (MODWT vs. DWT), filters (Daubechies Extremal Phase, Daubechies Least Asymmetric, and Coiflet families), and lengths (2--24) to determine their implications for the estimated values of network diagnostics. We quantify diagnostic variability, sensitivity, and utility in classifying resting state functional connectivity patterns extracted from people with schizophrenia and healthy controls using a previously-published fMRI data set \citep{bassett2012altered}. Our results demonstrate that wavelet method and length impact subsequent network diagnostics, but wavelet type has little effect. Based on our findings, we suggest that researchers use MODWT methods with a wavelet length of 8 or greater, and carefully report their choices to enhance comparability of results across studies.

\section{Materials and Methods}

\subsection{Ethics Statement}

All human subjects provied informed consent according to a protocol approved by the Institutional Review Board at the University of Minnesota.

\subsection{fMRI data acquisition and preprocessing}

Resting-state fMRI data from 29 healthy controls (11 females; age 41.1 $\pm$ 10.6 (SD)) and 29 participants with chronic schizophrenia (11 females; age 41.3 $\pm$ 9.3 (SD)) were included in this analysis (See \citep{camchong2011altered} for detailed characteristics of participants and imaging data). A Siemens Trio 3T scanner was used to collect the imaging data, including a 6-min (TR=2 secs; 180 volumes) resting-state fMRI scan, in which participants were asked to remain awake with their eyes closed, a field map scan, and a T1 MPRAGE whole brain volumetric scan. The fMRI data were preprocessed using FEAT (FMRIB's Software Library in FSL) with the following pipeline: deletion of the first 3 volumes to account for magnetization stabilization; motion correction using MCFLIRT; B0 fieldmap unwarping to correct for geometric distortion using acquired field map and PRELUDE+FUGUE52; slice-timing correction using Fourier-space time-series phase-shifting; non-brain removal using BET; regression against the 6 motion parameter time courses; registration of fMRI to standard space (Montreal Neurological Institute-152 brain); registration of fMRI to high resolution anatomical MRI; registration of high resolution anatomical MRI to standard space. Importantly, the two groups had similar mean RMS motion parameters: Two-sample t-tests of mean RMS translational and angular movement were both not significant ($p=0.14$ and $p=0.12$, respectively).

\subsection{Statistical analysis}
All calculations were done in MATLAB R2013b (The MathWorks Inc.). We used the WMTSA Wavelet Toolkit for MATLAB \url{(http://www.atmos.washington.edu/~wmtsa/)} to perform the wavelet decompositions, and we used the Brain Connectivity Toolbox \url{(https://sites.google.com/site/bctnet/)} to estimate values for network diagnostics.

\subsection{Network construction}

We extracted average time series for each participant from 90 of the 116 anatomical regions of interest (ROIs) defined by the AAL atlas \citep{Tzourio2002} covering the whole brain and including cortical and subcortical regions but excluding the cerebellar regions and vermis. We performed a battery of wavelet decompositions on each regional mean time series by varying wavelet method (DWT vs. MODWT), wavelet filter (Daubechies Extremal Phase, Daubechies Least Asymmetric, and Coiflet families), and wavelet length (2--24). In prior literature, both the discrete wavelet transform (DWT) and the maximal overlap discrete wavelet transform (MODWT) methods have been used to create functional connectivity matrices (see \citep{Deuker2009} and \citep{Vertes2012} respectively for examples). DWT is an orthogonal transform, just as the discrete Fourier transform (DFT); MODWT adds redundancy to DWT, and can be thought as a non-downsampled version of it \citep{percival2000wavelet}.

\begin{figure}[p]
	\centerline{\includegraphics[scale=0.6]{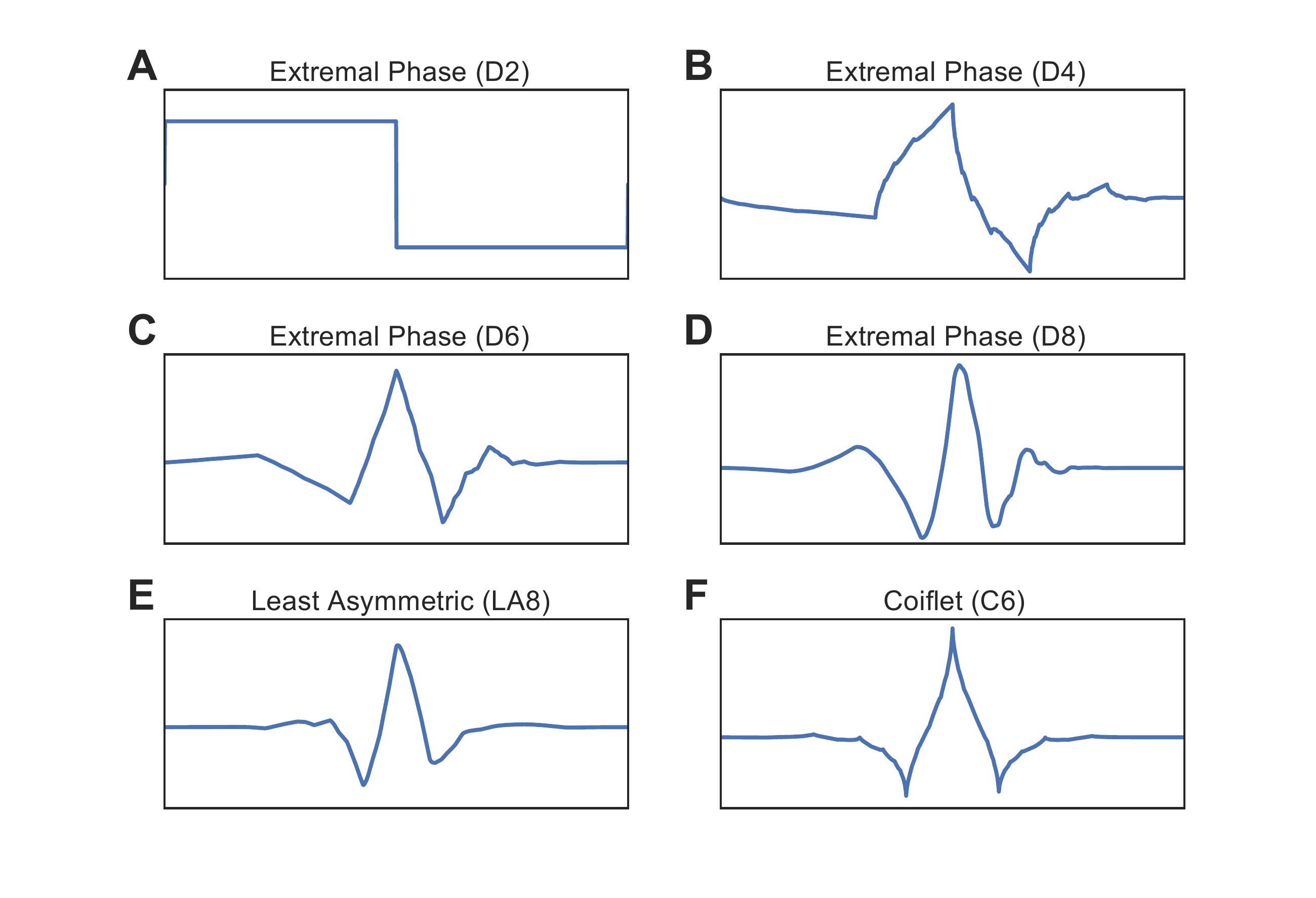}}
	\caption{\textbf{Example Wavelet Functions of Filters From Each Filter Type.} \emph{(A--D)} Daubechies Extremal Phase filter. \emph{(A)} Filter with length 2. \emph{(B)} Filter with length 4. \emph{(C)} Filter with length 6. \emph{(D)} Filter with length 8. \emph{(E)} Daubechies Least Asymmetric filter with length 8. \emph{(F)} Coiflet filter with length 6.}
	\label{fig1}
\end{figure}

Wavelet filter and length alter the symmetry and shape of the wavelet (see Fig.~\ref{fig1}). To examine the effect of wavelet filter, we apply Daubechies Extremal Phase, Daubechies Least Asymmetric, and Coiflet families \citep{percival2000wavelet}, which together constitute the most widely used orthogonal and compactly supported types of wavelet filters. We abbreviate these three filters types as D (Daubechies Extremal Phase), LA (Daubechies Least Asymmetric), and C (Coiflet). To examine the effect of wavelet length, we vary the length of the filter from 2 to 24. We refer to each wavelet type and length together; for example, \emph{D4} refers to the Daubechies Extremal Phase filter that has a length of 4.

Consistent with prior work \citep{Achard2006,Fornito2011}, we apply this battery of wavelet decompositions to each regional mean time series and extract wavelet coefficients for the first four wavelet scales, which in this case correspond to the frequency ranges 0.125$\sim$0.25 Hz (Scale 1), 0.06$\sim$0.125 Hz (Scale 2), 0.03$\sim$0.06 Hz (Scale 3), and 0.015$\sim$0.03 Hz (Scale 4). For each subject, wavelet method (DWT vs. MODWT), wavelet filter (Daubechies Extremal Phase, Daubechies Least Asymmetric, and Coiflet families), and wavelet length (2--24), we constructed a correlation matrix whose $ij^{th}$ elements were given by the estimated wavelet correlation between the wavelet coefficients of brain region $i$ and the wavelet coefficients of brain region $j$.

\subsection{Network diagnostics}

We characterized the organization of each functional connectivity matrix using both weighted and binary network diagnostics. To examine simple properties of the correlation matrix itself, we followed \citep{Lynall2010,bassett2012altered} and calculated (i) the mean correlation coefficient of the matrix as the average of the upper triangular elements of the matrix, and (ii) the variance of the correlation coefficients of the matrix as the variance of the upper triangular elements of the matrix.

To examine the topological properties of each functional connectivity matrix, we performed a cumulative thresholding approach \citep{bassett2012altered} by which we thresholded each matrix to maintain the strongest edges, giving a binary undirected network that has a density of 30\% (see the SI for examination of other thresholds). The choice of this threshold is based on a large and growing literature demonstrating small-world attributes of neuroimaging-based brain networks thresholded to retain this density \citep{Achard2007,Achard2006,Deuker2009,Lynall2010}. On this thresholded binary matrix, we calculated several network diagnostics, including the clustering coefficient, characteristic path length, global efficiency, local efficiency, modularity, and number of communities. See the Appendix for mathematical definitions of these diagnostics.

The maximization of modularity requires the investigator to make several methodological choices \citep{bassett2013robust}. Due to the heuristic nature of the Louvain algorithm \citep{blondel2008fast} used in maximizing the modularity quality function \citep{Newman2006} and the degeneracy of the modularity landscape \citep{Good2010}, we performed 20 optimizations of $Q$ for each functional connectivity matrix. The modularity values that we report are the mean values over these 20 optimizations. We also constructed a consensus partition \citep{lancichinetti2012consensus} from these optimizations using a method that compares the consistency of community assignments to that expected in a null model \citep{bassett2013robust}.

\subsection{Classification between healthy controls and schizophrenia patients}

To inform the utility of various wavelet methods, filters, and lengths in neuroimaging studies of functional brain network architecture, we performed a classification analysis in which we sought to classify functional connectivity matrices extracted from 29 healthy subjects from those extracted from 29 people with schizophrenia \citep{bassett2012altered}. This particular data set is well-suited to this study because it has been difficult to classify these two groups of subjects using binary networks constructed from traditional methods; the data set therefore offers a reasonable testbed for optimization of classification accuracy as a function of methodological variation. To perform this classification, we gathered all network diagnostics obtained in scale 2 (corresponding to the most commonly utilized frequency band for resting state network analyses \citep{Lynall2010,bassett2012altered}), and used a classification algorithm referred to as the C5.0 algorithm \url{(http://www.rulequest.com/see5-info.html)} to generate decision trees to classify data from healthy controls versus people with schizophrenia. The C5.0 algorithm supports boosting, and is faster and more memory efficient than the previous C4.5 algorithm \citep{quinlan1993c4}, which in turn is an extension of the earlier ID3 algorithm \citep{quinlan1986induction}. We generated decision trees with 10-trial boosting and 6-fold cross validation. The boosting method, AdaBoost, allows us to generate multiple decision trees for a given set of training data and combine them for better classification while avoiding overfitting \citep{freund1999short}. Utilizing the cross validation procedure, we randomly divided all of the subjects into 6 groups, and for each group, we trained a set of boosting decision trees on 5 groups and tested the decision trees on the remaining group. The results we report are the cumulative results across these 6 groups.

\section{Results}

In this section, we examine the effects of wavelet method (DWT vs. MODWT), wavelet filter (Daubechies Extremal Phase, Daubechies Least Asymmetric, and Coiflet families), and wavelet length (2--24) on (i) the estimated values of network diagnostics in healthy subjects, and (ii) the classification accuracy in distinguishing between functional connectivity matrices extracted from people with schizophrenia and healthy controls.

\begin{figure}[p]
	\centerline{\includegraphics[scale=0.5]{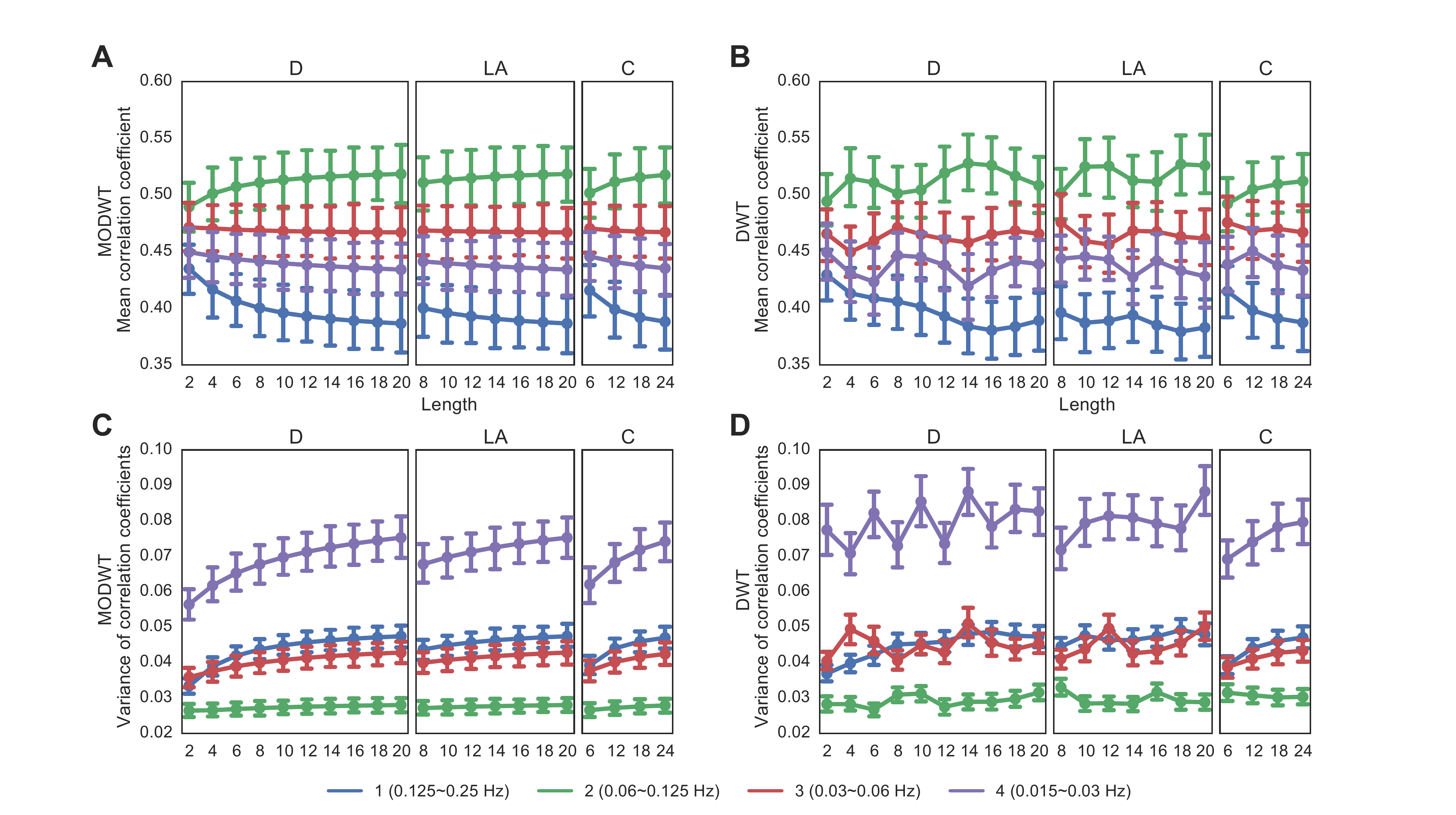}}
	\caption{\textbf{Effect of Wavelet Method on Mean and Variance of Correlation Coefficients} \emph{(A, B)} Mean correlation coefficients as a function of wavelet filter (Daubechies Extremal Phase, Daubechies Least Asymmetric, and Coiflet families) and wavelet length (2--24) observed when applying the \emph{(A)} MODWT and \emph{(B)} DWT. \emph{(C, D)} Variance of correlation coefficients as a function of wavelet filter (Daubechies Extremal Phase, Daubechies Least Asymmetric, and Coiflet families) and wavelet length (2--24) observed when applying the \emph{(C)} MODWT and \emph{(D)} DWT.  Wavelet scales are indicated by the color of the lines: scale 1 (approximately 0.125--0.25 Hz) is shown in blue, scale 2 (approximately 0.06--0.125 Hz) in green, scale 3 (approximately 0.03--0.06 Hz) in red, and scale 4 (approximately 0.015--0.03 Hz) in purple. Error bars indicate standard errors of the mean across 29 healthy subjects.}
	\label{fig2}
\end{figure}

\subsection{The Effect of Wavelet Method: DWT \emph{vs.} MODWT}

Both the DWT and the MODWT have previously been utilized to obtain wavelet coefficients for regional time series, prior to the construction of functional connectivity matrices representing graphs or networks (see \citep{Deuker2009} and \citep{Vertes2012} for recent examples). Here we performed a direct comparison between DWT and MODWT in terms of their effects on estimated network organization. In Fig.~\ref{fig2}, we show the mean and variance of the correlation coefficients of the functional connectivity matrices of healthy controls for all 3 wavelet filters, all 4 wavelet scales, and all wavelet lengths. In general, the shapes of the diagnostic versus wavelet length curves for both methods show qualitative similarities. We also observe that both DWT and MODWT give similar standard errors across subjects in the mean correlation coefficient and the variance of correlation coefficients.

Despite these gross qualitative similarities, we observe that the two methods differ in terms of (i) the variation of diagnostic values over wavelet lengths, and (ii) the magnitude of variance of correlation coefficients. Diagnostic values obtained using MODWT show a smooth change with increasing wavelet length, for all 3 wavelet filters and all 4 wavelet scales corresponding to different frequency bands (see Fig.~\ref{fig2} panels A and C). In contrast, diagnostic values obtained from DWT do not show smooth changes with increasing wavelet length (see Fig.~\ref{fig2} panels B and D). To quantify these observations, we calculated the sum of the absolute value of differences between diagnostics at consecutive lengths. For each scale and wavelet filter, we performed a paired $t$-test to test for differences in the mean. We found that -- indeed -- the variation of diagnostic values over wavelet lengths is significantly greater when using DWT than when using MODWT for all scales and all filters except scale 1 Coiflet; see Table~\ref{tab0}.

\begin{table}[p]
\centerline{
\begin{tabular}{ccdcdc}
\toprule
\multirow{3}{*}{Scale} & \multirow{3}{*}{Filter type} & \multicolumn{2}{C{4cm}}{Mean correlation coefficient} & \multicolumn{2}{C{4.5cm}}{Variance of correlation coefficients} \\
\cmidrule(lr){3-4} \cmidrule(lr){5-6}
& & \multicolumn{1}{c}{$t$} & $p$ & \multicolumn{1}{c}{$t$} & $p$ \\
\midrule
\multirow{3}{*}{1} & D & -6.00 & \textcolor{red}{0.0000} & -3.93 & \textcolor{red}{0.0005} \\
& LA & -5.38 & \textcolor{red}{0.0000} & -5.97 & \textcolor{red}{0.0000} \\
& C & 0.94 & 0.3561 & 0.70 & 0.4920 \\
\multirow{3}{*}{2} & D & -9.45 & \textcolor{red}{0.0000} & -9.98 & \textcolor{red}{0.0000} \\
& LA & -7.55 & \textcolor{red}{0.0000} & -7.64 & \textcolor{red}{0.0000} \\
& C & -3.32 & \textcolor{red}{0.0025} & -2.47 & \textcolor{red}{0.0200} \\
\multirow{3}{*}{3} & D & -6.98 & \textcolor{red}{0.0000} & -7.36 & \textcolor{red}{0.0000} \\
& LA & -10.01 & \textcolor{red}{0.0000} & -8.52 & \textcolor{red}{0.0000} \\
& C & -6.57 & \textcolor{red}{0.0000} & -5.51 & \textcolor{red}{0.0000} \\
\multirow{3}{*}{4} & D & -8.83 & \textcolor{red}{0.0000} & -9.89 & \textcolor{red}{0.0000} \\
& LA & -10.21 & \textcolor{red}{0.0000} & -8.44 & \textcolor{red}{0.0000} \\
& C & -9.99 & \textcolor{red}{0.0000} & -5.32 & \textcolor{red}{0.0000} \\			
\bottomrule
\end{tabular}
}
\caption{\textbf{Variation of Diagnostic Values Over Wavelet Lengths} $t$-values and $p$-values for two-sample $t$-tests measuring the differences in the sum of the absolute value of differences between diagnostics at consecutive lengths obtained from the MODWT approach as opposed to the DWT approach ($df$=28 over the 29 healthy control subjects). Paired $t$-tests were performed separately for each filter type (``D'' = Daubechies Extremal Phase, ``LA'' = Daubechies Least Asymmetric, and ``C'' = Coiflet) for each wavelet scale separately.}
\label{tab0}
\end{table}

Furthermore, the variance of correlation coefficients extracted using the MODWT method are smaller in magnitude than the variance of the correlation coefficients extracted using the DWT method (compare Fig.~\ref{fig2} panels C and D). To quantify this observation, we averaged the variance of the correlation coefficients over all wavelet lengths and filter types, separately for each scale. We performed a paired two-sided $t$-test to measure the difference between the average variance of correlation coefficients obtained using the DWT method versus those obtained using the MODWT method. We found that the average variance of the correlation coefficients was larger in the DWT case than in the MODWT case for all 4 wavelet scales: $t=5.87$ and $p<0.0001$ (Scale 1), $t=8.89$ and $p<0.0001$ (Scale 2), $t=14.64$ and $p<0.0001$ (Scale 3), and $t=9.44$ and $p<0.0001$ (Scale 4). Together these results are consistent with the theoretical notion that DWT provides more noisy estimates of structure than MODWT, and support the common preference in neuroimaging studies to use MODWT over DWT \citep{Achard2006}.

Based on its reliable variation with wavelet length, we restrict ourselves to the study of network diagnostics extracted using the MODWT method for the remainder of this paper.

\begin{figure}[p]
	\centerline{\includegraphics[scale=0.5]{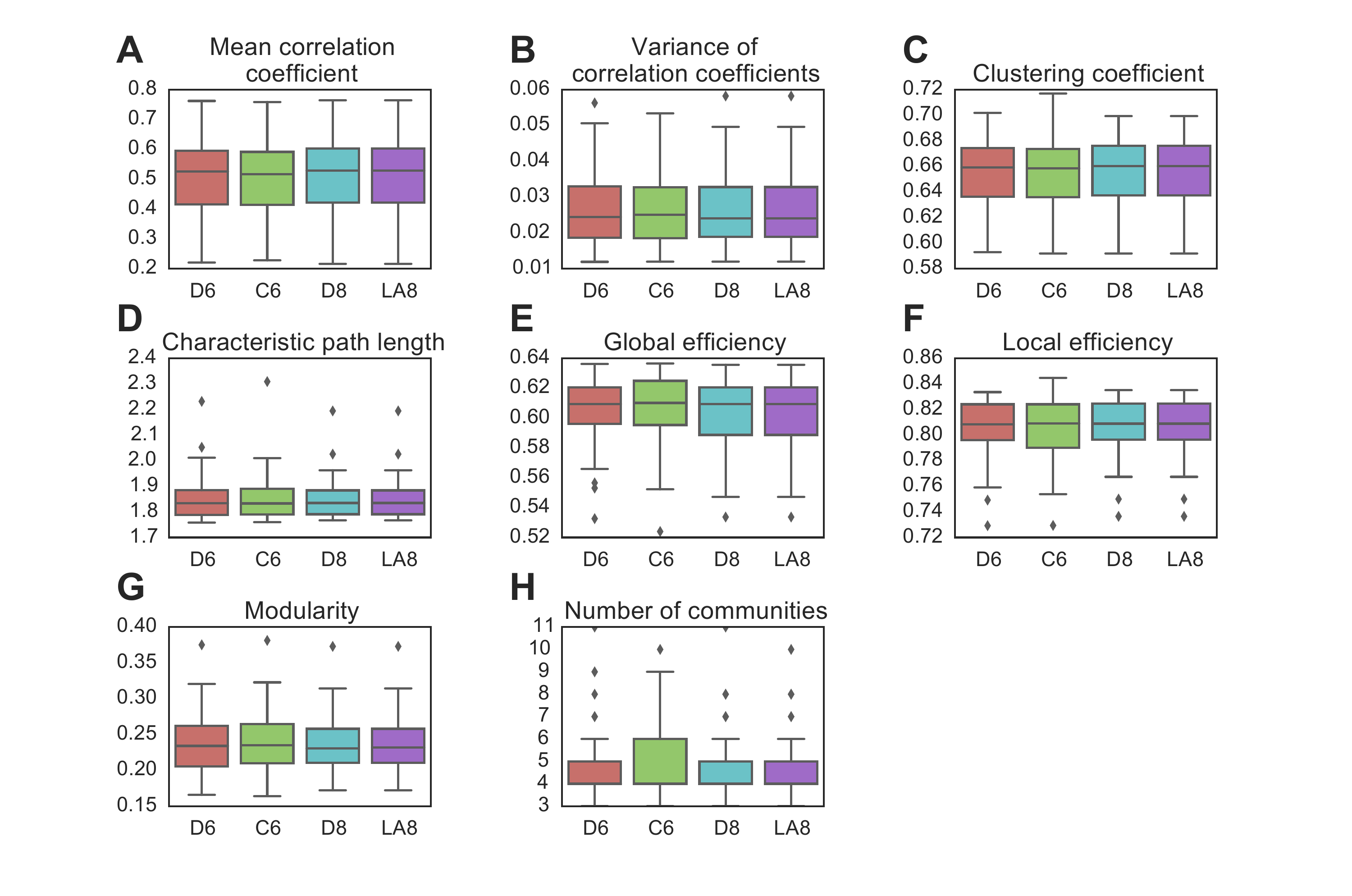}}
	\caption{\textbf{Effect of Wavelet Filter on Network Diagnostics} in wavelet scale 2 between pairs of wavelet filters with the same length. \emph{(A, B)} Weighted network diagnostics including \emph{(A)} mean correlation coefficient and \emph{(B)} variance of correlation coefficients. \emph{(C--F)} Binary network diagnostics calculated at a graph density of 30\% obtained through a cumulative thresholding procedure, including \emph{(C)} the clustering coefficient, \emph{(D)} characteristic path length, \emph{(E)} global efficiency, \emph{(F)} local efficiency, \emph{(G)} modularity index $Q$, and \emph{(H)} the number of communities. Boxplots indicate the median and quartiles of the data acquired from 29 health subjects. See Supplemental Materials for qualitatively similar results obtained at different scales and graph densities.}
	\label{fig3}
\end{figure}

\subsection{The Effect of Wavelet Filter Type}

In prior literature, many wavelet filters have been applied to the extraction of regional time series prior to functional brain network construction, including Daubechies \citep{Deuker2009}, and Least Asymmetric families \citep{Jakab2013}. Moreover, Coiflet wavelets have been shown to provide superior compression performance in magnetic resonance images \citep{Abu1999}. Here we performed a direct comparison between Daubechies Extremal Phase, Daubechies Least Asymmetric, and Coiflet families in terms of their effects on estimated network organization. To isolate the effect of wavelet filter, we examine network diagnostics obtained using each filter family and a fixed wavelet length. In Fig.~\ref{fig3}, we show representative results from a comparison of D6 and C6, and a comparison of D8 and LA8 in wavelet scale 2.  Qualitatively, we observe no significant differences in network diagnostics estimated from different wavelet filters of the same wavelet length. To confirm this observation quantitatively, we use a sign test (due to the skewed distribution of the data) to test the hypothesis that the difference median is zero between the distributions of diagnostics for D6 and C6, and between the distributions of diagnostics for D8 and LA8. Consistent with our qualitative observations, we find no significant differences (as defined as $p<0.05$ corrected for multiple comparisons using a conservative family-wise error correction). Note that we observe qualitatively similar results for other wavelet scales and other graph densities (see Supplemental Materials).

\begin{figure}[p]
	\centerline{\includegraphics[scale=0.45]{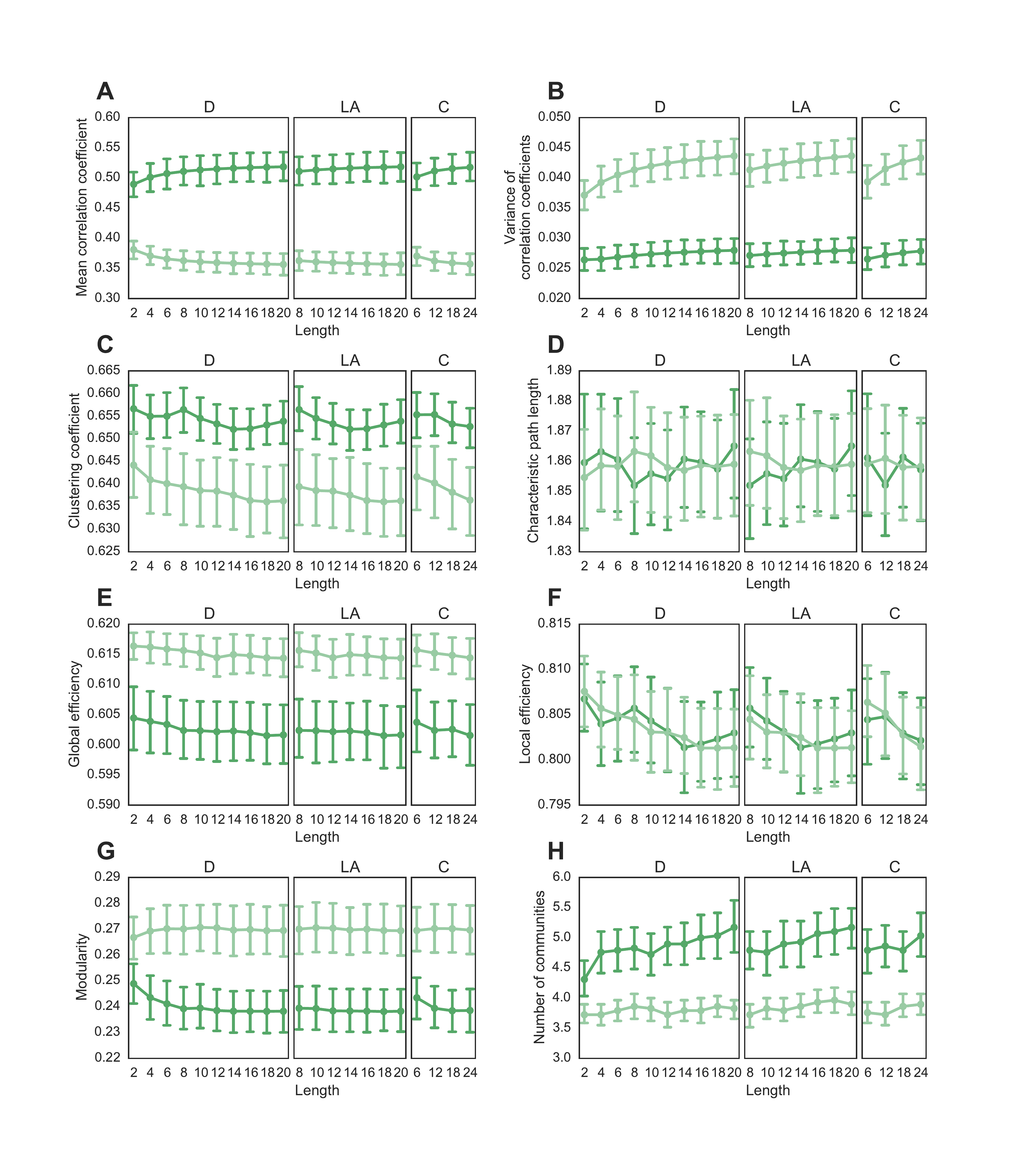}}
	\caption{\textbf{Effect of Wavelet Length on Network Diagnostics} in wavelet scale 2 for all wavelet filters. \emph{(A, B)} Weighted network diagnostics including \emph{(A)} mean correlation coefficient and \emph{(B)} variance of correlation coefficients. \emph{(C--F)} Binary network diagnostics calculated at a graph density of 30\% obtained through a cumulative thresholding procedure, including \emph{(C)} the clustering coefficient, \emph{(D)} characteristic path length, \emph{(E)} global efficiency, \emph{(F)} local efficiency, \emph{(G)} modularity index $Q$, and \emph{(H)} the number of communities.  The more saturated curves represent data from the 29 healthy controls, while the less saturated curves represent data from 29 people with schizophrenia. Error bars depict standard errors of the mean across subjects.  See Supplemental Materials for qualitatively similar results obtained at different wavelet scales.}
	\label{fig4}
\end{figure}

\subsection{The Effect of Wavelet Filter Length}

In prior literature, many wavelet lengths have been applied to the extraction of regional time series prior to functional brain network construction (for example see \citep{Deuker2009} and \citep{Jakab2013}). Here we performed a direct comparison between wavelet lengths 2 through 20 (Daubechies Extremal Phase), 8 to 20 (Daubechies Least Asymmetric), and 6 to 24 (Coiflet). Note these length choices were dictated by those available in the WMTSA toolbox (see Methods). Consistent with effects shown in Fig.~\ref{fig2}, we observe that the length of the wavelet filter affects network diagnostics differently; some diagnostics are affected significantly (such as the modularity index), and other diagnostics are affected very little (such as the characteristic path length); see Fig.~\ref{fig4}.

To quantify the differential sensitivity of network diagnostics to wavelet length, we performed a set of repeated measures ANOVA, for each diagnostic and each type of wavelet filter. Here, wavelet filter length was treated as a categorical factor, and diagnostic type was treated as a repeated measure. For complete results for each of these ANOVAs, see Table~\ref{tab1}. We observe that the mean and variance of correlation coefficients are significantly affected by wavelet length in all 3 wavelet filters. The characteristic path length and global efficiency are not significantly affected by wavelet length in any of the 3 wavelet filters. The clustering coefficient, local efficiency, modularity, and number of communities are affected by wavelet length in some but not all of the wavelet filters. These results demonstrate that network diagnostics are differentially sensitive to wavelet length, challenging the potential performance of meta-analyses that incorporate results obtained using different wavelet length and filters.

Note that we observe qualitatively similar results for other wavelet scales (see Supplemental Materials).

\begin{table}[p]
\centerline{
\begin{tabular}{>{\centering\arraybackslash}p{3cm}dcdcdc}
\toprule
& \multicolumn{2}{C{4cm}}{Daubechies Extremal Phase (dF=9,252)} & \multicolumn{2}{C{4cm}}{Daubechies Least Asymmetric (dF=6,168)} & \multicolumn{2}{C{3cm}}{Coif{}let (dF=3,84)} \\
\cmidrule(lr){2-3} \cmidrule(lr){4-5} \cmidrule(lr){6-7}
& \multicolumn{1}{c}{$F$} & $p$ & \multicolumn{1}{c}{$F$} & $p$ & \multicolumn{1}{c}{$F$} & $p$ \\
\midrule
Mean correlation coefficient & 14.63 & \textcolor{red}{0.0000} & 16.12 & \textcolor{red}{0.0000} & 16.29 & \textcolor{red}{0.0000} \\
Variance of correlation coefficients & 4.57 & \textcolor{red}{0.0000} & 13.28 & \textcolor{red}{0.0000} & 8.62 & \textcolor{red}{0.0000} \\
Clustering coefficient & 1.54 & 0.1333 & 4.47 & \textcolor{red}{0.0003} & 1.61 & 0.1937 \\
Characteristic path length & 0.37 & 0.9506 & 0.70 & 0.6503 & 0.53 & 0.6599 \\
Global efficiency & 0.95 & 0.4837 & 0.39 & 0.8852 & 1.18 & 0.3224 \\
Local efficiency & 1.60 & 0.1168 & 4.46 & \textcolor{red}{0.0003} & 1.32 & 0.2747 \\
Modularity & 6.88 & \textcolor{red}{0.0000} & 1.20 & 0.3089 & 5.07 & \textcolor{red}{0.0028} \\
Number of communities & 3.98 & \textcolor{red}{0.0001} & 3.41 & \textcolor{red}{0.0033} & 1.02 & 0.3898 \\
\bottomrule
\end{tabular}
}
\caption{\textbf{Effect of Wavelet Length.} Results of Repeated Measures ANOVAs for network diagnostics extracted from 29 healthy controls at scale 2 and a graph density of 30\%; wavelet length is treated as a factor and network diagnostic is treated as a repeated measure, separately for each wavelet filter type. Effects that are significant at $p<0.05$, uncorrected, are shown in red.}
\label{tab1}
\end{table}

\subsection{Classification in Psychiatric Disease}

Finally, we asked whether different wavelet filters provide different degrees of statistical sensitivity or classification accuracy when seeking to distinguish between functional connectivity matrices extracted from healthy controls versus those extracted from people with schizophrenia.

\begin{figure}[p]
	\centerline{\includegraphics[scale=0.5]{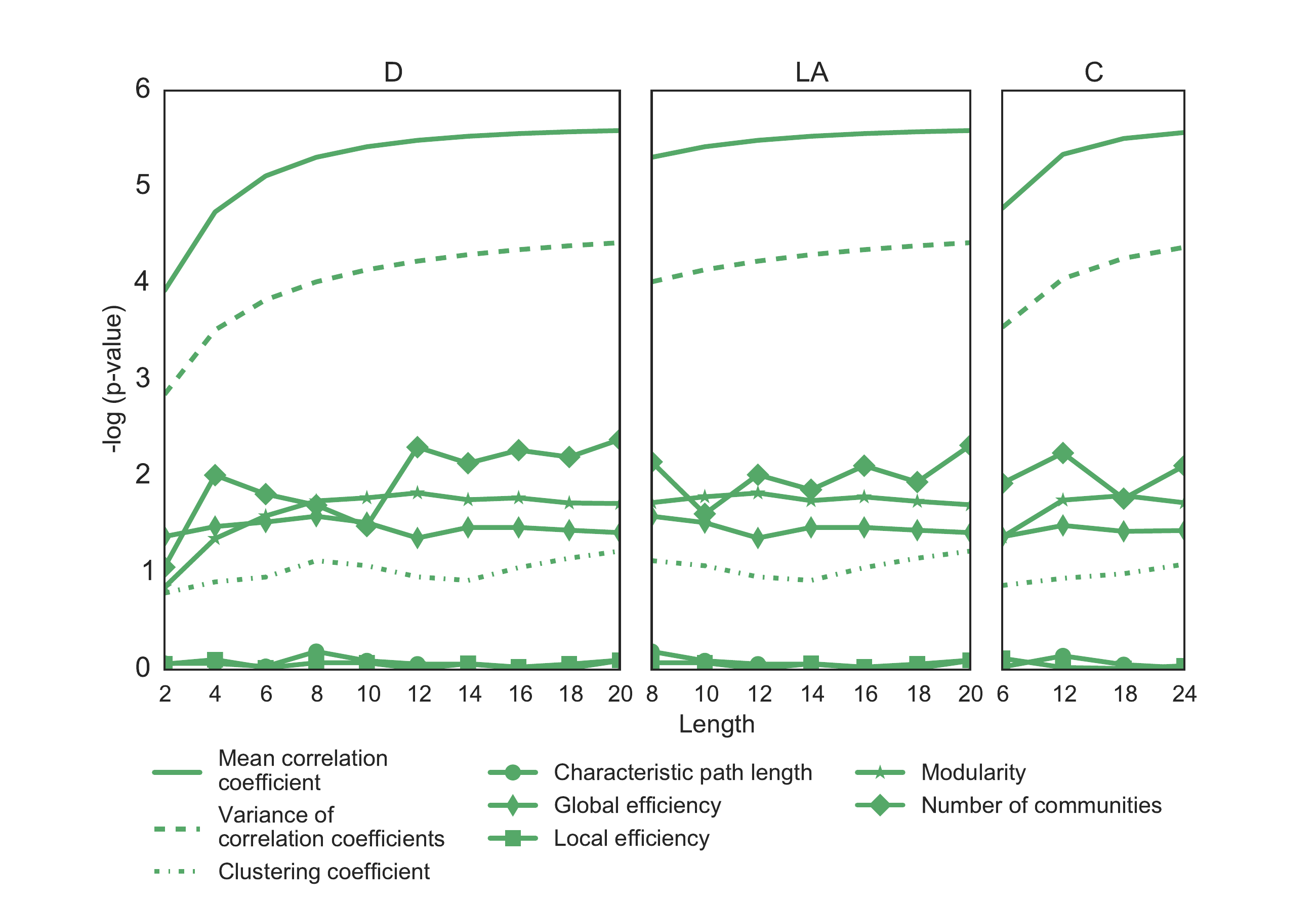}}
	\caption{\textbf{Effect of Wavelet Filter Type and Length on Statistical Sensitivity in Group Comparisons.} Negative common logarithm of the $p$-values obtained from two-sample $t$-tests between diagnostic values extracted from healthy control networks versus those extracted from schizophrenia patient networks. Higher values indicate greater group differences and lower values indicate weaker group differences. Network diagnostics are calculated for wavelet scale 2; for results in wavelet scale 1, see the SI.}
	\label{fig5_2}
\end{figure}

To determine whether different wavelet filters provide different degrees of statistical sensitivity for group comparisons, we first visually inspect diagnostic values in wavelet scale 2 as a function of filter type and length (compare dark and light lines in Fig.~\ref{fig4}). We observe that group differences in mean correlation coefficient, variance of correlation coefficients, clustering coefficient, modularity, and number of communities appear to be larger for longer wavelet lengths, across all three filter types. To quantify these observations, we performed a two-sample $t$-test between diagnostic values extracted from the two groups (patients vs. controls) for each filter type and length (see Fig.~\ref{fig5_2}). In general, we observe that the $p$-values decreased with increasing wavelet length (as demonstrated by the increase in the minus log $p$-values in Fig.~\ref{fig5_2}), suggesting that longer wavelets display greater statistical sensitivity to group differences in these data.  This finding was particularly salient for the mean correlation coefficient, variance of the correlation coefficients, clustering coefficient, modularity and number of communities, consistent with our visual inspection of Fig.~\ref{fig4}.

In the SI, we explore the dependence of these results on methodological choices in network construction including the measure of functional connectivity (partial correlation, wavelet coherence, and the wavelet correlation used in the main manuscript), strength of edges (strongest versus weakest \citep{bassett2012altered,schwarz2011negative}), and time series (wavelet details vs. wavelet coefficients). We observe that the effect of wavelet length is more salient (i) when using wavelet correlation than when using wavelet coherence or partial correlation, and (ii) when using the strongest 30\% connections or 10\% weakest connections than when using the 30\% or 1\% weakest connections. Results are consistent across the use of both wavelet details and wavelet coefficients. Based on prior work \citep{bassett2012altered}, we speculate that the networks constructed from the 1\% weakest connections display significant spatial localization and the networks that constructed from the 30\% weakest connections display significant random structure, together overshadowing the potential effects of wavelet length on group differences.

\begin{figure}[p]
	\centerline{\includegraphics[scale=0.5]{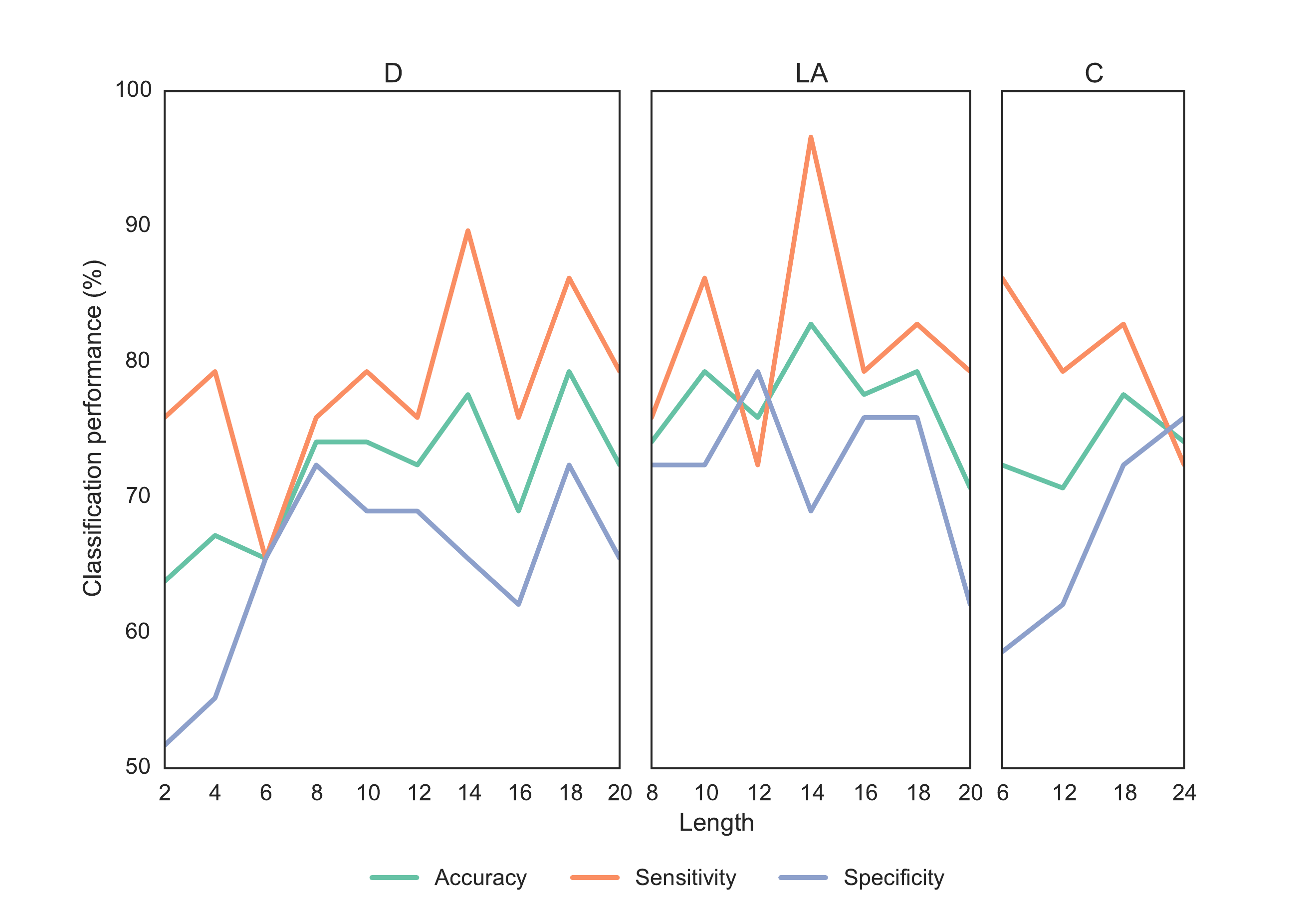}}
	\caption{\textbf{Effect of Wavelet Filter Type and Length on Classification.} Classification accuracy, sensitivity, and specificity as a function of wavelet filter type and length. Results are based on decision trees (see Methods) and distinguish between healthy controls and people with schizophrenia based on network diagnostics computed in wavelet scale 2. Note that we have regarded schizophrenia as positive, which clarifies the direction of the sensitivity and specificity estimates.}
	\label{fig6}
\end{figure}

We build on the above results drawn from parametric $t$-tests by applying non-parametric machine learning techniques to determine whether different wavelet filters provide different degrees of classification accuracy. Specifically, we generated decision trees (see Methods) to classify healthy controls and people with schizophrenia based on network diagnostics extracted from functional brain networks constructed from correlations in scale 2 wavelet coefficients. We observe that the classification accuracy ranged from approximately 63.8\% to approximately 82.8\%, the classification sensitivity ranged from approximately 65.5\% to approximately 96.6\%, and the classification specificity ranged from approximately 51.7\% to 79.3\% (see Fig.~\ref{fig6}). The poorest classification accuracy and specificity occurred in \emph{short} wavelets using the Daubechies Extremal Phase filter, and the best classification results occurred for relatively \emph{long} wavelets using the Daubechies Least Asymmetric filter (LA14), which gave 82.8\% accuracy and 96.6\% sensitivity. These results support those obtained from the parametric $t$-test analysis, that larger wavelet lengths display greater statistical sensitivity to group differences in these data.

\section{Discussion}

Wavelet-based methods offer extensive benefits in time series analysis and functional brain network construction. These include denoising capabilities \citep{Fadili2004}, robustness to outliers \citep{Achard2006}, utility in null model construction \citep{Breakspear2004}, frequency-specificity without edge effects \citep{percival2000wavelet}, and accurate estimates of functional connectivity in long memory processes \citep{Whitcher2000,Gencay2001}, such as those observed in fMRI time series \citep{Maxim2005,Wink2006,Achard2007,Bullmore2004,Achard2008}. Yet despite their utility, fundamental principles to guide the performance of wavelet-based methods remain largely undefined, hampering comparability and reproducibility of wavelet-based functional connectivity studies. Here we explicitly fill this gap by exploring the use of different wavelet methods (MODWT vs. DWT), filters (Daubechies Extremal Phase, Daubechies Least Asymmetric, and Coiflet families), and lengths (2--24) and by determining their implications for the estimated values of functional network diagnostics and the sensitivity to group differences. We found that the MODWT produces less variable estimates than the DWT method, and that wavelet length significantly impacts network diagnostic values and sensitivity to group differences. Collectively, our results underscore the importance of reporting the choices utilized in neuroimaging studies and provide concrete recommendations for these choices in wavelet-based analyses.

In the remainder of this section, we translate our results into concrete recommendations for the field, and we close with a brief discussion of important future directions.

\paragraph{The Choice of Wavelet Method}

The superior performance of MODWT in the context of the numerical experiments performed here is consistent with features of its theoretical construction \citep{whitcher2000wavelet}. First, and perhaps most importantly, MODWT is well defined for any signal length, making it statistically appropriate for the processing of arbitrary signals. In contrast, strictly speaking a DWT of level $J_0$ can be applied only to signals whose length is a multiple of $2^{J_0}$, significantly limiting its application to signals of arbitrary lengths.\footnote{In practice when applying the DWT to signals of arbitrary lengths, one can choose to avoid this issue -- as we did in this study -- by preserving at most one extra scaling coefficient at each level of wavelet decomposition.} Second, while DWT is an orthogonal transform, MODWT is not. In fact, MODWT is highly redundant and invariant under `circular shift' \citep{whitcher2000wavelet,percival2000wavelet}. This feature of MODWT preserves the smooth time-varying structure in regional time series that is otherwise lost during the application of DWT. In the context of human neuroimaging, analyses based on MODWT therefore more accurately reflect the dynamics of brain activity.

\paragraph{The Choice of Wavelet Filter Type and Length}

Wavelet filter types offer differently shaped wavelets that can be applied to empirical time series in a wavelet decomposition. While there is a generally well-accepted notion that one should choose a wavelet that displays similar time-varying features to the time series at hand, we observed that wavelet filter type had very little influence on network diagnostics extracted from resting state fMRI signals. The much larger factor impacting network diagnostics was the wavelet length, which tunes the fine-scale detail of the wavelet shape: larger wavelet length provides smoother wavelets. In general, network diagnostics obtained using the Daubechies Extremal Phase wavelets changed more from wavelet lengths 2 to 6 than from lengths 6 to 20. These results are intuitive: the changes in wavelet smoothness are more apparent at shorter wavelet lengths than at larger wavelet lengths, and their impact on estimated wavelet coefficients should follow. From a reliability perspective, we would argue that one would wish to choose a wavelet of a relatively larger length, to ensure that one's results are (i) not sensitive to artifacts of jagged edges in the wavelet and (ii) are relatively robust to small perturbations in wavelet length. Yet, it is important to keep in mind that very large wavelet lengths may suffer from the following limitations: (i) more coefficients may be influenced by boundary conditions, (ii) a decrease in the degree of localization of the wavelet coefficients, (iii) an increase in computational burden \citep{percival2000wavelet} . The ideal choice may therefore be a moderate length that retains the advantages of long wavelets without gaining any associated disadvantages.

\paragraph{Wavelets for Classification}

In our methodological recommendations thus far, we have called on arguments of reliability, insensitivity to artifact, and decreased variability to support specific choices in wavelet-based functional network analysis. In a final analysis we further asked whether one can support these choices based on differential sensitivity to group differences in functional network architecture. In analyses based on scale 2 wavelet coefficients (corresponding to 0.06--0.125 Hz), the answer is clear: longer wavelet lengths provide increased sensitivity to group differences as measured both by parametric $t$-tests and non-parametric machine learning algorithms based on decision trees. Using these longer wavelets, we observe significantly greater classification accuracy, sensitivity, and specificity values than those previously observed in this same data set \citep{bassett2012altered}, complementing prior work demonstrating differences in spontaneous low-frequency (\textless0.1 Hz) fluctuations in BOLD signal \citep{bluhm2007spontaneous,fornito2010can} and functional or structural network architecture \citep{lynall2010functional,liu2008disrupted,liang2006widespread,skudlarski2010brain} between schizophrenia patients and healthy controls. Thus, in addition to their benefits in terms of sensitivity and robustness, longer wavelets offer greater sensitivity to group differences in this data set, supporting their choice in the performance of wavelet-based analyses of resting state fMRI data more broadly. We speculate that there might be some underlying structural difference between the two groups of subjects that is consistent among individuals, and that the longer wavelet lengths smooth small differences between individuals so that large-scale differences are clearer. More generally, we speculate that larger wavelet lengths are better able to distinguish group-level features, while shorter wavelets may better distinguish individual-level features.

\paragraph{Methodological Considerations}

In general, our results point to the optimality of longer wavelets for functional network construction from low-frequency spontaneous fluctuations of the BOLD signal. However, it is interesting to note that for higher frequencies such as those probed by scale 1 coefficients (corresponding to 0.125--0.25 Hz), shorter wavelet lengths appear to provide better sensitivity to group differences; see the SI.  The assessment of these higher frequencies is uncommon in functional network construction, due to their decreased power and relative lack of structured topological architecture \citep{achard2006resilient}. Nevertheless, these results suggest that the optimal methodological choice for wavelet length might depend on the frequency band of interest, and therefore the properties of the signal being studied, an observation that might be particularly relevant in the assessment of functional networks in EEG and MEG data. Such a conclusion is supported by work identifying a variety of wavelet lengths and types as optimal for classification schemes in EEG signals \citep{subasi2007eeg} and other complex systems \citep{palit2010classification,semler2005wavelet}. More work is therefore necessary to determine rules of thumb for wavelet analysis that are generalizable across frequency bands and imaging modalities.

We have exercised these methods on functional networks constructed using the AAL atlas applied to resting state fMRI data, which represent common choices in functional network analysis in both health and disease. It will be interesting in future to assess the utility of these methods in other parcellation schemes and in task-based data.

\paragraph{Future Directions}

As a final note, it is worth pointing out that the wavelet decompositions utilized here build on procedures currently employed in the literature on functional brain network construction in an effort to provide the field with a few useful rules of thumb. However, other wavelet-based analysis techniques do exist -- including wavelet packets, dual-tree complex wavelet transforms, and double-density DWT -- that have not yet been applied to this problem, and it is not yet known whether these alternative techniques might provide complementary insights into whole-brain patterns of functional connectivity. It will be interesting in future to assess the utility of these alternative methods in reliably quantifying brain network organization and its alteration in disease states.

\appendix

\section{Relationship Between Sampling Frequency and Wavelet Scales}

The frequency ranges extracted by a wavelet decomposition directly depend on the sampling frequency of the data. It is therefore important to delineate which features of our results are generalizable across data sets acquired with different sampling frequencies. The data used here was acquired with a TR of $2$ s (a common choice), and therefore contains information up through the frequency $0.25$ Hz. A wavelet decomposition of this signal affects consecutive scales in which the observed signal is repeatedly convolved with a wavelet filter (which behaves as a high-pass filter) and a related scaling filter (which behaves as a low-pass filter). The first four scales therefore correspond to the frequency ranges of approximately $0.125-0.25$ Hz, $0.06-0.125$ Hz, $0.03-0.06$ Hz, and $0.015-0.03$ Hz, respectively. We note that different sampling frequencies may be used in other experiments, and the applicability of our specific results will depend on the degree of overlap in the frequency ranges of wavelet scales. However, our approach and conclusions regarding (i) the benefits of MODWT, (ii) the utility of moderate wavelet lengths, and (iii) the relatively small effect of wavelet filter are expected to be more generally applicable.

\section{Definitions of Network Diagnostics}

\begin{enumerate}
\item Clustering coefficient $C$: The clustering coefficient is used to quantify the local clustering properties of the network. First, the local clustering coefficient $C_i$ of a node $i$ can be defined as the fraction of actual edges between its neighbors \citep{watts1998collective}:
$$C_i=\frac{\Sigma_{j\neq h}A_{ij}A_{ih}A_{jh}}{k_i(k_i-1)},$$
where $\mathbf{A}$ refers to the adjacency matrix, and $k_i$ refers to the degree of node $i$. Then, the clustering coefficient of the network is defined as the mean of $C_i$ over all nodes.

\item Characteristic path length $L$: The characteristic path length is defined as the length of the geodesic path between two vertices, averaged over all pairs of connected vertices:
$$L=\frac{\Sigma_m\Sigma_{ij\in\mathcal{V}_m}d_{ij}}{\Sigma_mn_m^2},$$
where $\mathcal{V}_m$ refers to the set of vertices in connected component $m$, $d_{ij}$ refers to the geodesic distance between node $i$ and $j$, and $n_m$ refers to the number of nodes in connected component $m$.

\item Global efficiency $E_{\mathrm{glob}}$ \citep{latora2001efficient}: The global efficiency has been interpreted as a measure of how effectively information can be exchanged through the network. It is defined as follows:
$$E_{\mathrm{glob}}=\frac{1}{n(n-1)}\Sigma_{i\neq j}d_{ij}^{-1},$$
where $n$ is the number of nodes in the network.

\item Local efficiency $E_{\mathrm{loc}}$ \citep{latora2001efficient}: The local efficiency of node $i$ assesses the efficiency of the subgraph formed by the neighbors of $i$:
$$E_{\mathrm{loc},i}=\frac{\Sigma_{j\neq h}A_{ij}A_{ih}d_{jh}^{-1}}{k_i(k_i-1)}.$$
The local efficiency of the entire network is taken as the mean of $E_{\mathrm{loc},i}$ over all nodes in the network.

\item Modularity $Q$ \citep{newman2004finding,newman2004fast,newman2006mod}: The modularity of a network under a specific partitioning paradigm measures how well the network is divided into non-overlapping groups (or communities) of nodes such that the number of within-group edges is larger than expected in some null model \citep{newman2004finding,newman2004fast,newman2006mod,porter2009,fortunato2010}.  The modularity index is defined as:
$$Q=\Sigma_{ij}(A_{ij}-\frac{k_ik_j}{2l})\delta(c_i,c_j),$$
where $l$ is the number of edges in the network, $c_i$ and $c_j$ are the communities containing nodes $i$ and $j$, respectively, and $\delta(c_i,c_j)$ is the Kronecker delta. In this study, we presented the maximum modularity value obtained with the Louvain algorithm \citep{blondel2008fast} over 100 nearly degenerate solutions \citep{Good2010}.
\end{enumerate}

\paragraph{Acknowledgments}
ZZ acknowledges support from the US-China Summer Research Program of the University of Pennsylvania. This work was supported by the John D. and Catherine T. MacArthur Foundation, the Alfred P. Sloan Foundation, the Army Research Laboratory and the Army Research Office through contract numbers W911NF-10-2-0022 and W911NF-14-1-0679, the National Institute of Mental Health (2-R01-DC-009209-11), the National Institute of Child Health and Human Development (1R01HD086888-01), the Office of Naval Research, and the National Science Foundation (BCS-1441502 and BCS-1430087). The funders had no role in study design, data collection and analysis, decision to publish, or preparation of the manuscript. We thank Sarah Feldt Muldoon for helpful comments on an early version of the manuscript.

\bibliographystyle{elsarticle-harv}
\bibliography{wavelet_paper}





\end{document}